\documentclass[12pt,twoside]{article}   
\usepackage[super,sort&compress,comma]{natbib}

\usepackage{fancyhdr}		

\newcommand{\captionv}[3]{\begin{center}\parbox{#1cm}{\caption[#2]{{\sf #3}}}
        \end{center}}

\usepackage[section]{placeins}   %

\usepackage{graphicx}
\usepackage{subfigure}

\makeatletter \renewcommand\@biblabel[1]{$^{#1}$} \makeatother
\newcommand{\cen}[1]{\begin{center} #1 \end{center}}

\newcommand{\vy}{\boldsymbol{y}}
\newcommand{\pred}{\hat{\boldsymbol{y}}}
\newcommand{\sens}{\eta_{\text{sens}}}
\newcommand{\spec}{\eta_{\text{spec}}}

\newcommand{\modified}[1]{\textcolor{black}{#1}}

\usepackage[mathlines]{lineno}
\usepackage{amsmath}
\usepackage{makecell}
\usepackage{multirow}

\usepackage{hyperref}
\hypersetup{ colorlinks,
    citecolor=blue,
    filecolor=blue,
    linkcolor=blue,
    urlcolor=blue
}

\usepackage{xcolor}

\definecolor{gray}{rgb}{0.6,0.6,0.6}
\definecolor{red}{rgb}{0.85,0,0}
\definecolor{green}{rgb}{0,0.85,0}
\definecolor{blue}{rgb}{0,0,0.85}
\definecolor{beige}{rgb}{0.92,0.87,0.78}
\usepackage[all]{hypcap}    

\begin{document}

\cen{\sf {\Large {\bfseries Deep learning for brain metastasis detection and segmentation in longitudinal MRI data } \\  
\vspace*{10mm}
Yixing Huang$^{1,2}$, Christoph Bert$^{1,2}$, Philipp Sommer$^{1,2}$,\\Benjamin Frey$^{1,2}$, Udo Gaipl$^{1,2}$, Luitpold V. Distel$^{1,2}$,\\ Thomas Weissmann$^{1,2}$,  Michael Uder$^5$, Manuel A. Schmidt$^3$,\\Arnd D\"orfler$^3$, Andreas Maier$^4$, Rainer Fietkau$^{1,2}$, Florian Putz$^{1,2}$} \\
1. Department of Radiation Oncology, Universit\"atsklinikum Erlangen, Friedrich-Alexander-Universit\"at Erlangen-N\"urnberg (FAU), Erlangen, Germany \\
2. Comprehensive Cancer Center Erlangen-EMN (CCC ER-EMN), Erlangen, Germany.\\
3. Department of Neuroradiology, Universit\"atsklinikum Erlangen, FAU, Erlangen, Germany\\
4. Pattern Recognition Lab, FAU, Erlangen, Germany\\
5. Institute of Radiology, Universit\"atsklinikum Erlangen, FAU, Erlangen, Germany
\vspace{5mm}\\
Version typeset \today\\
}

\pagenumbering{roman}
\setcounter{page}{1}
\pagestyle{plain}
Author to whom correspondence should be addressed. email: yixing.yh.huang@fau.de; florian.putz@uk-erlangen.de \\

The implementation for this work is available to public: 

\url{https://github.com/YixingHuang/DeepMedicPlus}

\begin{abstract}
\noindent {\bf Purpose:} Brain metastases occur frequently in patients with metastatic cancer. Early and accurate detection of brain metastases is essential for treatment planning and prognosis in radiation therapy. Due to their tiny sizes and relatively low contrast, small brain metastases are very difficult to detect manually. With the recent development of deep learning technologies, several researchers have reported promising results in automated brain metastasis detection. However, the detection sensitivity is still not high enough for tiny brain metastases, and integration into clinical practice in regard to differentiating true metastases from false positives is challenging.\\
{\bf Methods:} The DeepMedic network with the binary cross-entropy (BCE) loss is used as our baseline method. To improve brain metastasis detection performance, a custom detection loss called volume-level sensitivity-specificity (VSS) is proposed, which rates metastasis detection sensitivity and specificity at a (sub-)volume level. 
As sensitivity and precision are always a trade-off, either a high sensitivity or a high precision can be achieved for brain metastasis detection by adjusting the weights in the VSS loss without decline in dice score coefficient for segmented metastases. To reduce metastasis-like structures being detected as false positive metastases, a temporal prior volume is proposed as an additional input of DeepMedic. The modified network is called DeepMedic+ for distinction. Combining a high sensitivity VSS loss and a high specificity loss for DeepMedic+, the majority of true positive metastases are confirmed with high specificity, while additional metastases candidates in each patient are marked with high sensitivity for detailed expert evaluation. \\
{\bf Results:} Our proposed VSS loss improves the sensitivity of brain metastasis detection, increasing the sensitivity from 85.3\% for DeepMedic with BCE to 97.5\% for DeepMedic with VSS. Alternatively, the precision is improved from 69.1\% for DeepMedic with BCE to 98.7\% for DeepMedic with VSS. Comparing DeepMedic+ with DeepMedic with the same VSS loss, 44.4\% of the false positive metastases are reduced in the high sensitivity model and the precision reaches 99.6\% for the high specificity model. The mean dice coefficient for all metastases is about 0.81. With the ensemble of the high sensitivity and high specificity models, on average only 1.5 false positive metastases per patient need further check, while the majority of true positive metastases are confirmed. \\
{\bf Conclusions:} Our proposed VSS loss and temporal prior improve brain metastasis detection sensitivity and precision. The ensemble learning is able to distinguish high confidence true positive metastases from metastases candidates that require special expert review or further follow-up, being particularly well-fit to the requirements of expert support in real clinical practice. This facilitates metastasis detection and segmentation for neuroradiologists in diagnostic and radiation oncologists in therapeutic clinical applications.
\end{abstract}

\newpage     

\tableofcontents

\newpage

\setlength{\baselineskip}{0.7cm}      

\pagenumbering{arabic}
\setcounter{page}{1}
\pagestyle{fancy}

\section{Introduction}
Patients with metastatic cancer have a high risk of developing brain metastases (BM), with an approximate incident rate up to 40\% \cite{tabouret2012recent}. The progression of BM leads to reduced or absent efficacy of common systemic treatments. Therefore, successful treatment of BM is very crucial for patient survival and quality of life \cite{steinmann2009effects,le2021eano}. As whole-brain radiotherapy causes cognitive impairments, stereotactic radiosurgery (SRS) has gained increasing preference for BM treatment \cite{chang2009neurocognition,kocher2014stereotactic,brown2016effect,sperduto2020beyond}. SRS delivers high focused radiation to metastasis regions with little dose to surrounding normal brain tissues. Hence, it causes much fewer side effects compared with whole-brain radiotherapy. For SRS treatment planning, the number, size, boundary, and location of BM are essential information, which requires accurate detection and subsequent segmentation of BM. Currently, BM are identified manually by neuroradiologists and radiation oncologists, which is time-consuming and suffers from inter-observer variability \cite{charron2018automatic,xue2020deep,growcott2020inter}. Especially, small metastases are easily overlooked in manual detection \cite{xue2020deep}, as they are located only in a few image slices and typically have low contrast. In addition, some anatomical structures, such as blood vessels, appear very similar to BM in 2D intersection planes, which makes manual identification challenging \cite{kocher2020applications}. Although the omitted small metastases have a better chance to be identified in the follow-up scans when they grow larger, treatment at later rather than the earliest stages will negatively affect patient prognosis, as larger metastases are more difficult to control and may be associated with distressing symptoms \cite{baschnagel2013tumor}. Therefore, computer assisted automated BM identification has important clinical value.

For automated BM detection and segmentation, conventional machine learning methods such as template matching \cite{perez2016brain,sunwoo2017computer}, support vector machine \cite{gonella2019investigating} and AdaBoost \cite{zhang2020deep} have been applied. However, they have been proven inferior to the latest deep learning methods \cite{gonella2019investigating,zhang2020deep}. With the recent explosion of deep learning techniques, although the majority of researchers focus on the segmentation of primary brain tumors like gliomas \cite{bakas2018identifying}, the research of deep learning for BM detection and segmentation is growing \cite{cho2021brain}. Considering neural network architectures for BM segmentation, 3D U-Net \cite{hu2019multimodal,lu2019automated,bousabarah2020deep,xue2020deep} and DeepMedic \cite{kamnitsas2017efficient,liu2017deep,lu2019automated,charron2018automatic,hu2019multimodal} are the most common two networks. Other neural networks include GoogLeNet \cite{grovik2020deep}, V-Net \cite{gonella2019investigating}, Faster R-CNN \cite{zhang2020deep}, single-shot detectors \cite{zhou2020computer}, and custom convolutional neural networks (CNNs) \cite{losch2015detection,dikici2020automated}. To save memory, the GoogLeNet approach \cite{grovik2020deep} utilizes seven slices, one central slice plus six neighboring slices, as a 2.5D model. All other neural networks use 3D subvolumes for training.

To train segmentation neural networks, loss functions of binary cross-entropy (BCE), dice similarity coefficient (DSC), and intersection over union (IOU) are generally applied. To improve BM segmentation performance, new loss functions are proposed. For example, Bousabarah et al. \cite{bousabarah2020deep} proposed a soft dice loss computed from three spatial scales, which are the outputs of three down-sampled layers from the 3D U-Net. As dice coefficients can be dominated by large tumors, the contribution of small metastasis regions will be underweighted. To overcome this problem, Hu et al. \cite{hu2019multimodal} proposed a volume-aware dice loss, where the dice coefficients are reweighted by metastasis volume sizes. However, all the above methods have limited performance, either in detection sensitivity or precision. For example, the volume-aware dice loss together with ensemble learning only achieves a precision of 0.79 \cite{hu2019multimodal}; the soft dice loss only achieves a sensitivity of 0.82 \cite{bousabarah2020deep}; the custom CNN \cite{losch2015detection} and the GooLeNet \cite{grovik2020deep} both achieve a sensitivity of 0.83. Although Dikici et al. \cite{dikici2020automated} and Charron et al. \cite{charron2018automatic} have achieved relatively high sensitivity of 0.9 and 0.93 respectively, they both have high average false positive (FP) rates of 9.1 and 7.8 FP metastases per patient, respectively.
 Especially, their detection sensitivity of tiny metastases is not satisfactory. For example, the GoogLeNet-based method \cite{grovik2020deep} only achieves 50\% sensitivity for metastases smaller than 7\,mm \cite{kocher2020applications}; The two-step U-Net method \cite{xue2020deep}, achieving 100\% accuracy,  is only evaluated on BM larger than 0.07\,cm$^3$ with a median size of 2.22\,cm$^3$.
 
To adjust sensitivity and specificity, a sensitivity-specificity segmentation loss has been proposed by Brosch et al. on MRI brain images, but for the application of sclerosis lesion segmentation \cite{brosch2015deep}. This loss function is defined based on sensitivity-specificity error (SSE) in a voxel level and hence called SSE for short in this work. SSE has been applied to brain tumor segmentation. However, it is reported inferior to the weighted BCE loss \cite{sudre2017generalised}. For the specific application of BM identification, sensitivity-specificity based loss functions have not been investigated yet. In our work, SSE is found effective to tune the sensitivity and precision for BM identification. In clinical practice, the first primary goal is to correctly detect the existence of metastases with high sensitivity and specificity irrespective of the exact three-dimensional extent of each metastasis. With this purpose and independent from SSE, a new sensitivity-specificity based loss function with a novel formulation in a volume-level is proposed, which is demonstrated to have better performance than SSE in BM identification.

Among the above methods, \cite{charron2018automatic,hu2019multimodal,bousabarah2020deep,grovik2020deep} are based on multi-sequence MRI data, as complementary information from multi-sequence imaging is beneficial for BM detection and segmentation. However, T2-weighted, T2-FLAIR \cite{hajnal1992use}, and diffusion tensor imaging (DTI) \cite{le2001diffusion} images routinely available typically have low resolution, especially slice resolution, which limits their benefits in detecting tiny BM. In addition, the registration among multi-sequence data increases the requirements for standardized sequences and in clinical routine complete multi-sequence datasets are frequently not available. As single-modal high-resolution contrast-enhanced 3D T1-weighted (T1w) sequences are the most important sequences for BM identification in clinical practice and clinical research \cite{kaufmann2020consensus}, BM identification in single-modal data therefore has important clinical value.
In addition, patients with BM need follow-up scans to monitor BM radiotherapy treatment response and check for the development of new BM. As a result, longitudinal MRI volumes typically are available for each patient. For manual BM identification, clinicians easily tend to pay more attention to the regions where old BM exist. Therefore, there is a risk to ignore new BM, especially such new BM have tiny volume sizes and low contrast.
So far, the investigation on deep learning-based BM detection in longitudinal MRI data has not been studied yet. In this work, the application of deep learning for BM detection and segmentation in single-modal longitudinal MRI data is investigated.

\begin{figure}[t]
\centering
\includegraphics[width=1.0\linewidth]{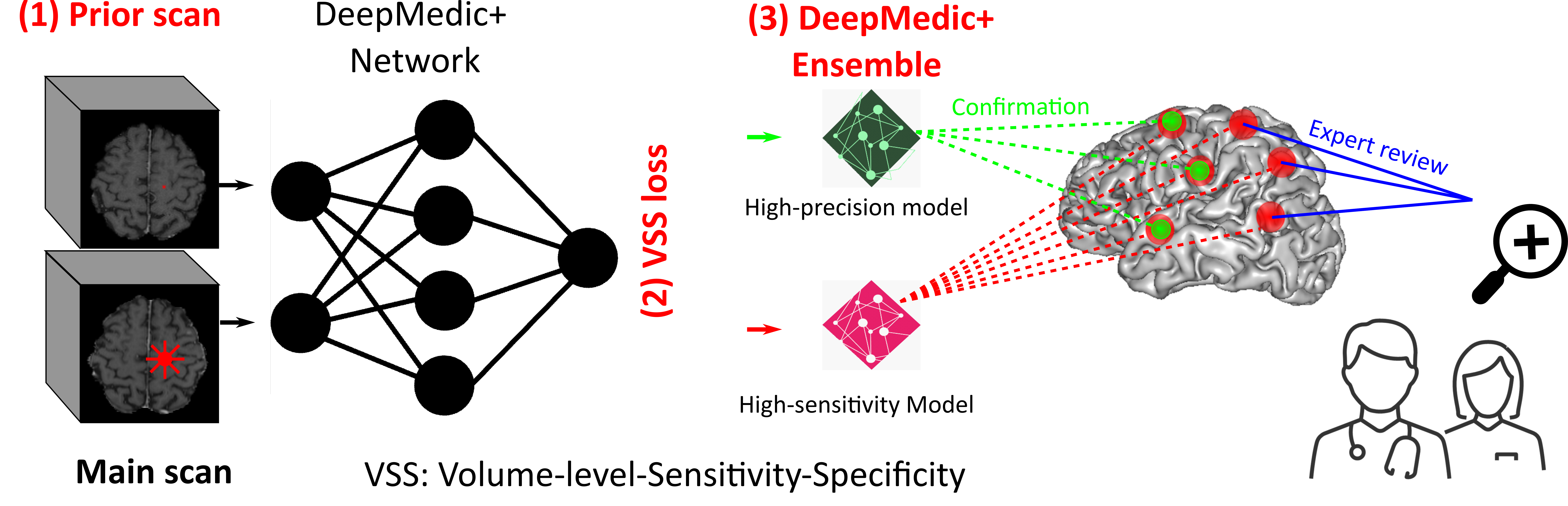}
\captionv{15}{}{\modified{A graphical summary of the three main contributions in this work.}}
\label{Fig:graphicalSummary}
\end{figure}

The contribution of this work lies mainly in the following aspects: a) A new loss function called volume-level sensitivity-specificity (VSS) loss, based on the definitions of sensitivity and specificity exactly at a (sub)volume level, is proposed to improve tiny metastases detection and to adjust sensitivity and specificity; b) To the best our knowledge, our work is the first to integrate temporal prior information for automated BM identification; c) An ensemble learning strategy that achieves high sensitivity for BM detection and distinguishes high-confidence metastases from metastases candidates, which require expert review or additional follow-up, to accommodate requirements of expert support in real clinical practice. \modified{A graphical summary of the above three main contributions is displayed in Fig.\,\ref{Fig:graphicalSummary}.}

\section{Materials And Methods}
\subsection{Baseline method: DeepMedic}
In this work, we apply the DeepMedic \cite{kamnitsas2017efficient} as our baseline method because of its efficacy in various brain tumor segmentation tasks. DeepMedic is originally proposed for general brain tumor segmentation. In brain tumor segmentation, especially for BM segmentation, class imbalance is a major issue as normal tissue voxels outnumber tumor voxels. To overcome this problem, DeepMedic samples volume segments online to keep class balance during training. In addition, multi-scale features are extracted via parallel convolutional pathways.


\subsection{Volume-level sensitivity-specificity (VSS) loss}

Conventional loss functions \cite{jadon2020survey,ma2021loss} like BCE, DSC, and IOU evaluate the output of a neural network at a voxel level for segmentation purpose. Such a segmentation needs to identify the existence, location, and three-dimensional extent of a metastasis, which is challenging in general. Optimal BM segmentation is particularly difficult to achieve with conventional voxel level loss functions, since the vast differences in volume sizes (or voxel numbers) among metastases leads to limited detection rate of tiny metastases and the partial volume effect results in large uncertainties in segmentation of tiny metastases when they approach individual voxel size. Instead, for clinical applications, algorithms are typically evaluated in a patient (volume) or lesion (metastasis) level. For example, sensitivity stands for the ability of a test to correctly identify patients with a disease, while specificity defines the ability of a test to correctly identify people without the disease \cite{swift2020sensitivity}. For BM segmentation in 3D volumes (same for subvolumes), the first primary clinical requirement is to correctly detect the existence of metastases with high sensitivity and specificity irrespective of the exact three-dimensional extent of each metastasis. If a metastasis exists in a volume (volume-level positive case), in case of a voxel-level classification task, this can be translated to maximize the probability of at least one predicted voxel being located in a true metastasis region (sensitivity); When no metastasis exists in a volume (volume-level negative case), the network needs to minimize the probability of any voxel being a false-positive metastasis (specificity).

\textbf{Sensitivity (positive case):} We denote the segmentation label of the $i$-th volume by $\vy_i$. Correspondingly, the prediction of the $i$-th volume is denoted by $\pred_i$. The label is binary, while the prediction is a probability between 0 and 1. With this definition, $\max(\vy_i)$ represents whether any metastasis exists in the $i$-th volume. If $\max(\vy_i) = 1$, there is at least one metastasis; otherwise, no metastasis exists. $\pred \cdot \vy_i$ is the intersection region, where $\cdot$ is voxel-wise multiplication. Then $\max(\pred_i \cdot \vy_i)$ represents the maximum voxel level probability that at least one metastasis has been detected within the $i$-th volume (probability from 0 to 1). Note that $\max(\pred_i \cdot \vy_i)$ instead of $\max(\pred_i) \cdot \max(\vy_i)$ is used to make sure the detected metastasis mask has overlap with the reference mask. In a batch with $B$ samples, $\sum_{i=1}^B\max(\pred_i \cdot \vy_i)$ stands for the number of correctly detected volumes (when the maximum probability approaches 1 in the ideal case). The total number of volumes containing metastases is $\sum_{i=1}^B\max(\vy_i)$. Accordingly, we can define the sensitivity as follows,

\begin{equation}
\sens = \frac{\sum_{i=1}^B\max(\pred_i \cdot \vy_i)}{\sum_{i=1}^B\max(\vy_i) + \epsilon},
\end{equation}
where $\epsilon$ is a small value to avoid division by zero. 

\textbf{Specificity (negative case):} In the negative case, $1 -\max(\vy_i)$ equals one for the $i$-th segmentation label, and $\sum_{i=1}^B(1 -\max(\vy_i))$ is the total number of negative volumes. Similarly, $1 - \max(\pred_i)$ equals one means that the neural network has predicted a tumor probability of zero for every voxel in the volume. The total number of volumes, which are correctly segmented as fully negative, is $\sum_{i=1}^B\left(\left(1 -\max(\vy_i)\right) \cdot \left(1 - \max\left(\pred_i\right)\right)\right)$. Accordingly, we can calculate the specificity as follows,

\begin{equation}
\spec = \frac{\sum_{i=1}^B\left(\left(1 -\max(\vy_i)\right) \cdot \left(1 - \max\left(\pred_i\right)\right)\right)}{\sum_{i=1}^B(1 -\max(\vy_i)) + \epsilon}.
\end{equation}

$\sens$ ensures that at least one metastasis can be detected in a volume if metastases exist, while $\spec$ ensures no FP metastasis presents in a fully negative volume. With the two metrics, a volume-level sensitivity-specificity (VSS) loss is defined,
\begin{equation}
\ell_{\text{VSS}} = 1 - (\alpha \cdot \sens + (1 - \alpha) \cdot \spec),
\label{eqn:MetDetLoss}
\end{equation}
where $\alpha$ ($0 \leq\alpha \leq 1$) is a parameter to adjust the weights of $\sens$ and $\spec$, as it is always a trade-off between sensitivity and specificity. According to its definition, the VSS loss focuses on the detection of metastases instead of segmentation, incorporating maximum predicted voxel-level probabilities in a continuous fashion. Because the sensitivity definition includes a overlap criterion and specificity is evaluated exclusively in volumes without metastases, high predicted tumor probabilities inside true metastases improve the cost function, whereas discrepancies between predictions and ground truth in the ambiguous periphery of metastases does not affect it for volumes harbouring metastases. Therefore, a conventional segmentation loss like BCE or DSC needs to be used together with the VSS loss so that the neural network is able to perform segmentation as well. In this work, we combine the VSS loss with the BCE loss as a default joint loss, denoted by JVSS,
\begin{equation}
\ell_{\text{JVSS}} = \ell_{\text{VSS}} + \ell_{\text{BCE}}.
\label{eqn:MetDetLoss}
\end{equation}

The SSE loss proposed for sclerosis lesion segmentation {was} defined as follows \cite{brosch2015deep},
\begin{equation}
SSE = \alpha \frac{\sum_p\left(\vy(p) - \hat{\vy}(p)\right)^2\vy(p)}{\sum_p\vy(p)} + (1-\alpha)\frac{\sum_p\left(\vy(p) - \hat{\vy}(p)\right)^2\left(1 - \vy(p)\right)}{\sum_p(1 - \vy(p))},
\end{equation}
where $p$ is the voxel index and hence the loss function is defined at a voxel-level as opposed to a subvolume level. The left part is sensitivity error and the right part is the specificity error. The parameter $\alpha$ adjusts the weights of these two parts. This loss function does not follow the definitions of sensitivity and specificity exactly as it focuses on the sensitivity and specificity errors. In this work, the performance comparison of our proposed sensitivity-specificity loss function with SSE in BM identification is investigated.

\subsection{Temporal prior path}
\begin{figure}[t]
\centering
\begin{minipage}{0.24\linewidth}
\subfigure[Prior image]{
\includegraphics[width=\linewidth]{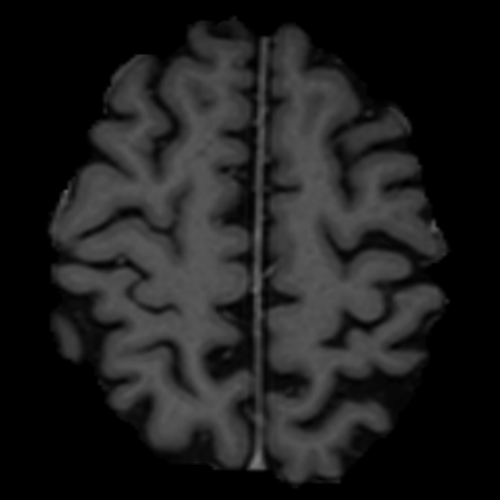}
\label{subfig:priorInput}
}
\end{minipage}
\begin{minipage}{0.24\linewidth}
\subfigure[Main image]{
\includegraphics[width=\linewidth]{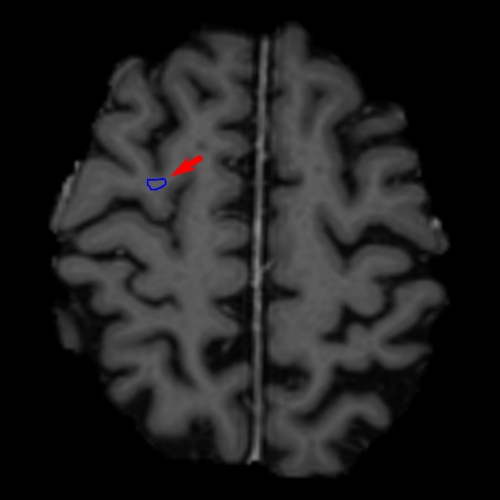}
\label{subfig:mainInput}
}
\end{minipage}
\begin{minipage}{0.24\linewidth}
\subfigure[Difference image]{
\includegraphics[width=\linewidth]{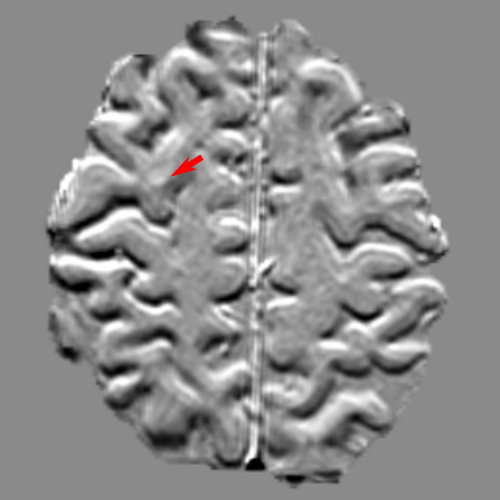}
\label{subfig:difference}
}
\end{minipage}
\captionv{15}{}{Exemplary images acquired from different time points: (a) temporal prior image; (b) current main image, where the contour indicated by the arrow is segmented incorrectly as a metastasis by a neural network that used the prior image as an additional channel as opposed to an additional pathway; (c) difference between (a) and (b), where the bright region indicated by the arrow is the main cause for the incorrect segmentation. }
\label{Fig:priorExemplaryImage}
\end{figure}
\begin{figure}[t]
\centering
\includegraphics[width=0.8\linewidth]{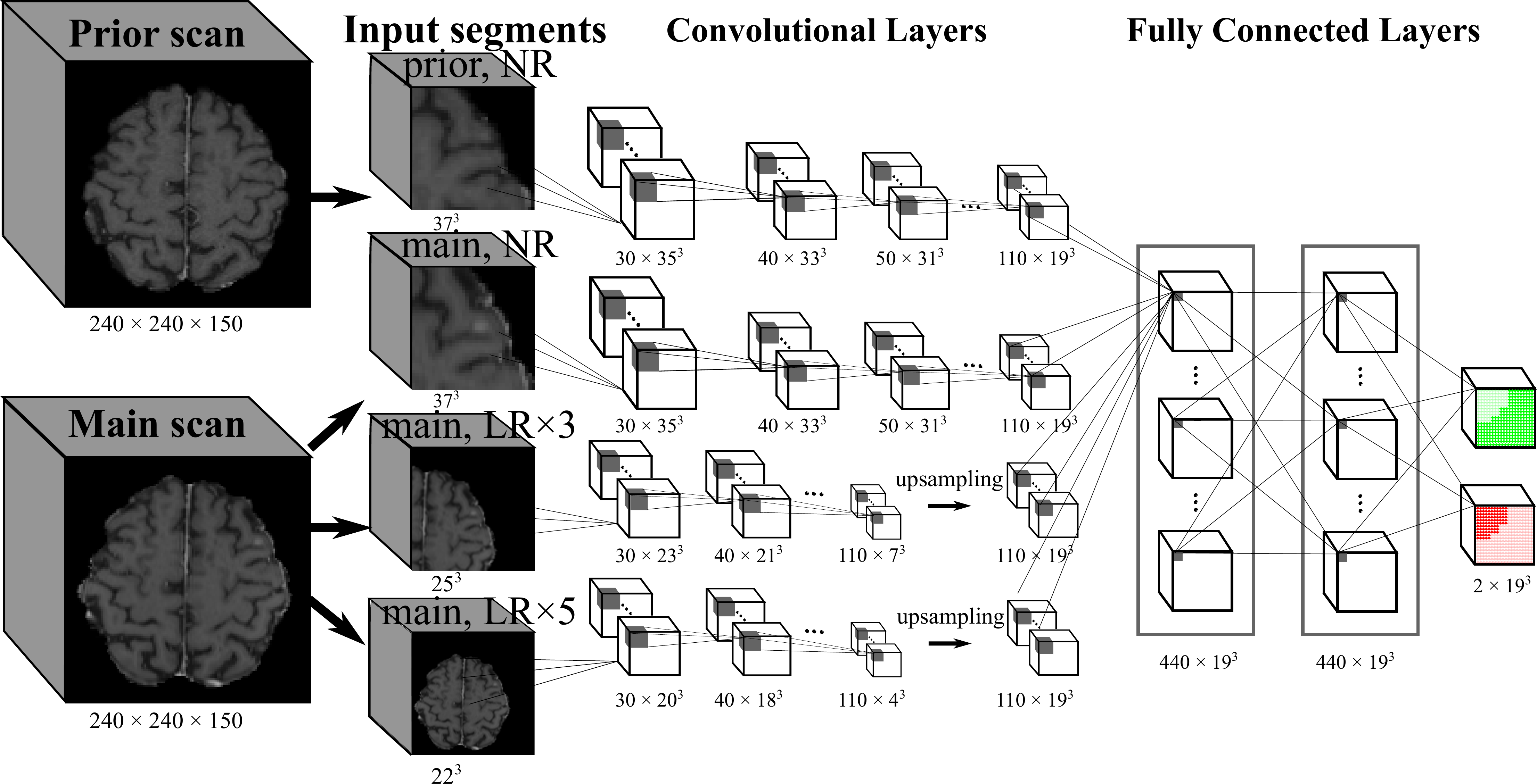}
\captionv{15}{}{The modified DeepMedic architecture, called DeepMedic+ in this work, uses one additional path for temporal prior volumes. The normal resolution (NR) subvolumes from both the prior and main datasets as well as two low resolution (LR) subvolumes from the main dataset are input segments of different network paths. The neurons of the last layers have receptive fields of $19^3$ voxels. \modified{The final two output channels sum to 1.} The feature map dimensions during training in our work are displayed as \textit{Number $\times$ Size}. Note that the feature dimensions are only for training. For inference, a different input size can be used, since DeepMedic is a fully convolutional network. Modified from \cite{kamnitsas2017efficient}.}
\label{Fig:modifiedDeepMedic}
\end{figure}

In contrast-enhanced MRI images with 3D volumetric gradient echo imaging (e.g., BRAVO, GE Healthcare; MPRAGE, Siemens Healthcare; 3D TFE; Philips Healthcare) \cite{tong2020advanced}, many structures such as blood vessels are also enhanced by contrast agents. Hence, they have similar appearance to metastases in the regard to intensity, shape, and size in 2D intersectional planes. As a consequence, it is very challenging for human experts to distinguish them. BM differ from blood vessels in several aspects. One is that although they appear similar in 2D intersectional planes, they have distinct morphological appearance in 3D space. For example, enhanced vascular structures are tube-like structures with bifurcations, while the majority of BM are sphere-like structures \cite{pope2018brain}. Therefore, 3D neural networks are advantageous over 2D networks to extract 3D features. In addition, BM have relatively larger morphological changes over time than normal tissues. Without treatment BM volume size typically grows faster than normal tissues due to the high proliferation rate of tumor cells. With treatment BM volume sizes change depending on treatment response, for example, volume size decreases in regression, increases in progression, and oscillates (increases first and later decreases) in pseudo-progression \cite{oft2020volumetric}. In radiation therapy, patients have regular follow-up MRI scans in approximately every 4-6 weeks. Therefore, by comparing two images acquired from two time points, if a high contrast structure emerges or grows, this structure has high confidence to be a metastasis. To integrate such temporal prior information in deep learning, two potential ways are possible: using the temporal prior volume as an additional input channel or as an additional input path. As anatomical structures imaged at different time points cannot be perfectly registered to the same position, using temporal prior volumes as an additional channel will result in a high FP rate. An example is displayed in Fig.\,\ref{Fig:priorExemplaryImage}, where a normal tissue region indicated by the arrow in Fig.\,\ref{subfig:mainInput} is segmented incorrectly as a metastasis. Such mis-segmentation is mainly caused by the imperfect registration, since high intensity difference is observed in the corresponding region of the difference image (Fig.\,\ref{subfig:difference}). To avoid such problem, in this work we propose to put the temporal prior volume as an additional input path, where features from two time points are merged at deep layers. For distinction, the modified DeepMedic architecture is called DeepMedic+ in this work. The overall DeepMedic+ architecture is displayed in Fig.\,\ref{Fig:modifiedDeepMedic}, where the normal resolution subvolumes from both the prior and main datasets as well as two low resolution subvolumes from the main dataset are fed into DeepMedic+. Note that for volumes without any temporal prior, an empty prior volume with zero values is used.

\subsection{Ensemble learning}
For segmentation outputs from neural networks trained with conventional losses like BCE, a threshold can be adjusted to get a balance between sensitivity and specificity. All the segmentation masks generated in this work use a default probability threshold of 0.5, which is a default setting of DeepMedic as well as a convention for segmentation applications. Since a large $\alpha$ value leads to high sensitivity while a small one leads to high specificity with our proposed JVSS loss. We can train the same network DeepMedic+ twice with different $\alpha$ values to get a high sensitivity model ($\mathcal{M}_\text{sens}$) and a high specificity model ($\mathcal{M}_\text{spec}$), respectively. The final segmentation mask is the union of the two segmentation masks from $\mathcal{M}_\text{sens}$ and $\mathcal{M}_\text{spec}$. The metastases predicted by $\mathcal{M}_\text{spec}$ have high confidence as true positive metastases,  
while $\mathcal{M}_\text{sens}$ predicts additional metastasis candidates with high sensitivity for detailed analysis by clinical experts or to undergo additional imaging studies or follow-up imaging.

\subsection{Experimental Setup}
\subsubsection{Dataset}
770 contrast enhanced T1 volumes from 176 patients using the MRI magnetization-prepared rapid gradient echo (MPRAGE) sequence \cite{brant1992mp} from a longitudinal study \cite{oft2020volumetric,putz2020fsrt} are used for evaluation. The MPRAGE MRI sequence is used in our regular clinical routine. Please find detailed information about data preprocessing and the distribution of BM in \cite{oft2020volumetric,putz2020fsrt}. The 770 volumes are in an alphabetical order according to patient names. Among them, first 600 volumes are used for training, 67 volumes for validation to monitor convergence and overfitting, and the last 103 volumes are used for test. Since the distribution of BM is independent from patient names, it is equivalent to select BM distributions randomly.
The training volumes are from 135 patients and 466 of them have temporal prior volumes. In total 1503 metastases are contained in the training dataset and among them 667 (44.4\%) have a volume size smaller than 0.1\,cm$^3$, which corresponds to a diameter of about 6\,mm.
The test volumes are from 32 patients (excluding any patient in the training datasets) and 71 of them have temporal prior volumes. In total, 278 metastases are contained in the test dataset. Among all the metastases, 130 (47\%) of metastases have a volume size smaller than 0.1\,cm$^3$. All volumes are resampled to an isotropic 1\,mm$^3$ resolution with dimensions $240 \times 240 \times 155$ and affinely registered. 
N4 bias correction \cite{tustison2010n4itk} is applied to get homogeneous intensity.

\subsubsection{Training and test parameters}
All models are trained on an NVIDIA Quadro RTX 8000 GPU with Intel Xeon Gold 6158R CPUs. Each training takes about 21 hours. Based on the validation loss, 60 epochs are used to train each model.
The initial learning rate is 0.001. The RMSProp optimizer \cite{tieleman2012lecture} and Nesterov momentum \cite{sutskever2013importance} with the moment value $m=0.6$, $\rho=0.9$ and $\epsilon = 10^{-4}$ are applied. 
The class-balance sampling with 50\% probability of tumor segments is applied to extract training samples. During training, the input segments for the main path have a size of $37 ^3$. As 9 convolutional layers with $3^3$ convolution kernels are used without padding, the receptive field is $(1+9\times 2)^3 = 19^3$ and the effective segment size in the network output is $(37 - 9 * 2)^3 = 19^3$. The segments are augmented with random intensity scaling, flipping and rotation. DeepMedic as well as DeepMedic+ is a fully convolutional network, which does not require fixed input size. For inference, to save computation on cropping input subvolumes, larger input segments with size $45^3$ are applied. The network configurations used in our work are recommended by the authors of DeepMedic \cite{kamnitsas2017efficient}, which can be found in their public GitHub repository\footnote{https://github.com/deepmedic/deepmedic}.

\subsubsection{Evaluation metrics}

For clinical applications, the number, location, and size of BM are very important for accurate treatment. Hence, our evaluation will focus on metastasis-level metrics. Since all the volumes contain BM, the metric of specificity is not used. Instead, the sensitivity and precision are calculated at a metastasis-level based on the number ($\#$) of TP, FP, and FN metastases. The average number of FP metastases per complete volume (i.e., per patient) is also reported. The metric of DSC can reflect the accuracy of location and size of segmented BM. In practice, BM segmentation is performed after detection confirmation. Therefore, the mean DSC (mDSC) of all TP metastases detected by a network model is calculated. The reference BM masks are segmented manually by experienced radiation oncologists.

\section{Results}

\subsection{Results of the VSS loss}

\begin{table}[h]
\begin{center}
\captionv{15}{}{Metastasis detection accuracy of DeepMedic with the JVSS loss using different $\alpha$ values.}
\label{Tab:MetDetLoss}
\begin{tabular}{|l|l|c|c|c|c|c|c}
\Xhline{3\arrayrulewidth}
Network& Loss & Sensitivity & Precision & $\#$FP & mDSC \\
\Xhline{3\arrayrulewidth}
\multirow{7}{*}{DeepMedic} &BCE & 0.853 & 0.691 & 106 & 0.789 \\
\cline{2-6}
& JVSS$_{\alpha =1}$ & 0.975  & 0.274 & 718 &0.808 \\
& JVSS$_{\alpha =0.995}$ &0.946  & 0.516 & 247 &0.798 \\
& JVSS$_{\alpha =0.99}$ & 0.914 & 0.736 & 91 &0.801 \\
& JVSS$_{\alpha =0.95}$ & 0.881 & 0.918 & 22  & 0.792\\
& JVSS$_{\alpha =0.9}$ & 0.860& 0.930 & 18  & 0.788 \\
& JVSS$_{\alpha =0.5}$ & 0.802 & 0.987 & 3  & 0.755 \\
\hline
\multirow{2}{*}{DeepMedic+}&JVSS$_{\alpha =0.995}$ & 0.932 & 0.621 & 158  &0.808 \\
& JVSS$_{\alpha =0.5}$ & 0.842 & 0.996 & 1 & 0.760 \\
\Xhline{3\arrayrulewidth}
\end{tabular}
\end{center}
\hspace{60pt}\small{ $\quad $ Number of total metastases: 278}
\vspace{10pt}
\end{table}

\begin{figure}[h!]
\centering
\begin{minipage}{0.24\linewidth}
\subfigure[]{
\includegraphics[width=\linewidth]{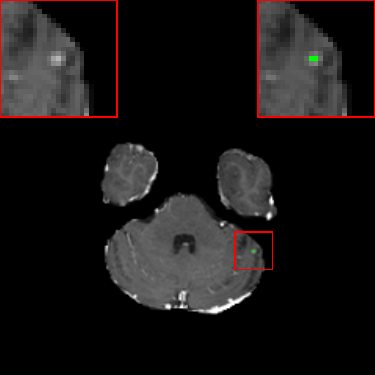}
\label{subfig:volumeS36RGB}
}
\end{minipage}
\begin{minipage}{0.24\linewidth}
\subfigure[]{
\includegraphics[width=\linewidth]{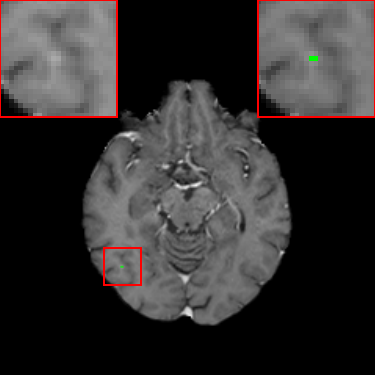}
\label{subfig:volumeS60RGB}
}
\end{minipage}
\begin{minipage}{0.24\linewidth}
\subfigure[]{
\includegraphics[width=\linewidth]{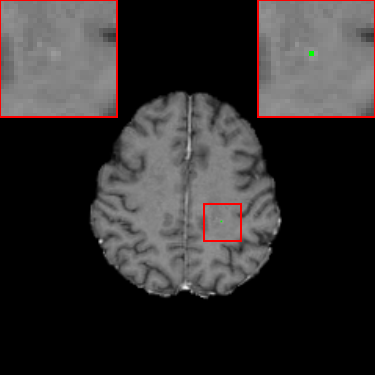}
\label{subfig:volumeS107RGB}
}
\end{minipage}
\captionv{15}{}{Three examples of FN metastases by $\mathcal{M}_\text{baseline}$ (the baseline DeepMedic model trained with BCE loss), which are detected by DeepMedic with the JVSS ($\alpha = 0.995$) loss. The segmentation masks are indicated by green color. Two zoomed-in ROIs without and with the segmentation mask are displayed on the left and right top corners respectively for each patient.}
\label{Fig:FalseNegativeExamples}
\end{figure}

The metastasis-level evaluation of DeepMedic and DeepMedic+ models trained with different loss functions is displayed in Tab.\,\ref{Tab:MetDetLoss}. The baseline DeepMedic model achieves a sensitivity of 0.853 (237/278) and a precision of 0.691 (237/(237 + 106)). In other words, among the 278 metastases, 237 TP metastases are detected, while 106 FP metastases are also marked. The FN metastases are mainly tiny metastases, as displayed in the top left regions-of-interest (ROIs) of Fig.\,\ref{Fig:FalseNegativeExamples} where three exemplary FN metastases are displayed.

With the JVSS loss, depending on the value of $\alpha$ in Eqn.\,(\ref{eqn:MetDetLoss}), the sensitivity and precision can be adjusted. When $\alpha = 1$ where high sensitivity is desired, DeepMedic achieves a high sensitivity of 0.975 (271/278) but a low precision of 0.275 (271/(271 + 718)). Among all the 278 metastases, 271 are successfully detected, but 718 FP metastases are detected. With a slight smaller $\alpha$ value, i.e., $\alpha = 0.995$, a high sensitivity of 0.946 (263/278) is achieved, while the precision is increased to 0.516 (263/(263 + 247)). The number of FP metastases is decreased from 718 to 247. In the top right ROIs of Fig.\,\ref{Fig:FalseNegativeExamples}, the metastasis positions of the three exemplary images are correctly detected by DeepMedic trained with the JVSS loss ($\alpha = 0.995$). 

\begin{figure}[t]
\centering
\begin{minipage}{0.26\linewidth}
\subfigure[]{
\includegraphics[width=\linewidth]{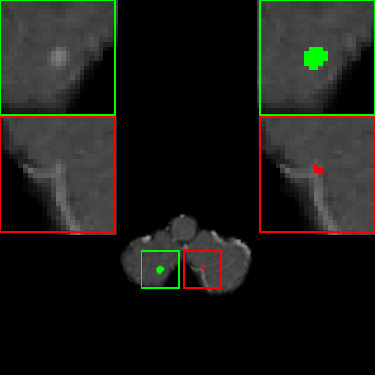}
\label{subfig:volumeS13RGB}
}
\end{minipage}
\begin{minipage}{0.26\linewidth}
\subfigure[]{
\includegraphics[width=\linewidth]{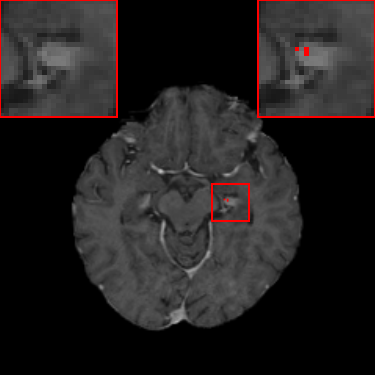}
\label{subfig:volumeS62RGB}
}
\end{minipage}
\begin{minipage}{0.26\linewidth}
\subfigure[]{
\includegraphics[width=\linewidth]{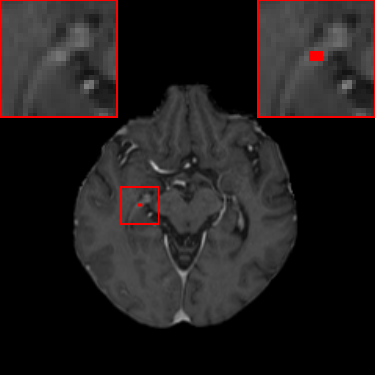}
\label{subfig:volumeS59RGB}
}
\end{minipage}
\captionv{15}{}{The three false positive metastases detected by DeepMedic with the JVSS ($\alpha = 0.5$) loss are marked by red color. The green region in (a) is a detected true positive metastasis.}
\label{Fig:FalsePositiveCases}
\end{figure}

When $\alpha$ is decreased in the JVSS loss, the sensitivity is decreased, but the precision is improved. When $\alpha = 0.9$, the sensitivity of DeepMedic with the JVSS loss is comparable to that of DeepMedic with BCE only. However, the precision is increased from 0.691 (237/(237 + 106)) to 0.930 (239/(239 + 18)). When $\alpha = 0.5$, DeepMedic achieves a high precision of 0.987, with only 3 FP metastases remaining. The three FP metastases are displayed in Fig.\,\ref{Fig:FalsePositiveCases}. In Fig.\,\ref{subfig:volumeS13RGB}, the green region is a TP metastasis and its boundary is segmented well. The red region in Fig.\,\ref{subfig:volumeS13RGB} is a FP metastasis, which is a connecting point of two vessels.  The right vesicle in Fig.\,\ref{subfig:volumeS62RGB} has higher intensity than the surrounding tissues and it has a sphere-like structure in the 3D space. That is why it is falsely detected as a metastasis. The marked region in Fig.\,\ref{subfig:volumeS59RGB} has relatively higher intensity than the surrounding vessels. Therefore, whether it is a false or true positive metastasis is controversial, if we look at one single volume only. After checking with the follow-up scans, this region does not grow at all and hence we confirm that it is a FP metastasis.
 
\begin{figure}
\centering
\includegraphics[width=0.6 \linewidth]{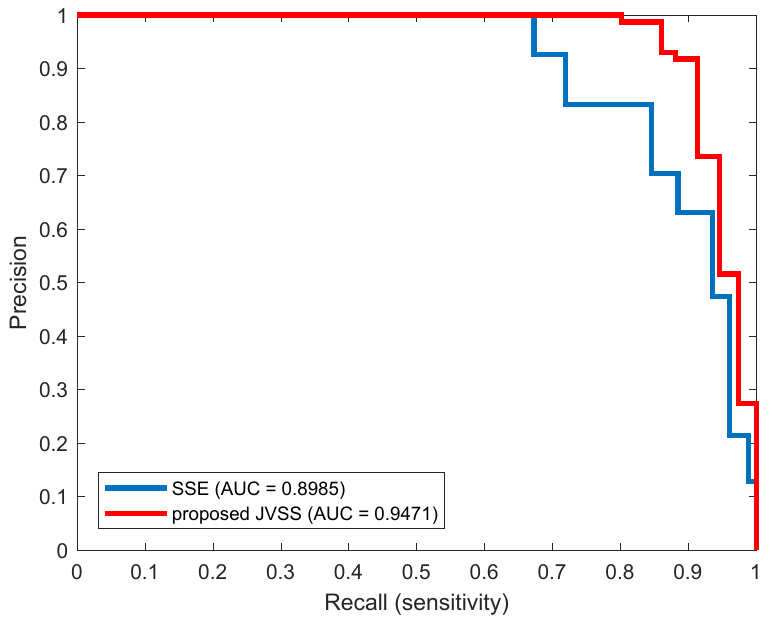}
\captionv{15}{}{The precision-recall curves of DeepMedic with our proposed VSS loss and the SSE loss by adjusting the weight $\alpha$.}
\label{Fig:SSEcomparison}
\end{figure}

The metastasis detection accuracy of DeepMedic with SSE \cite{brosch2015deep} is also evaluated. For a better comparison between SSE and our proposed VSS, their precision-recall (sensitivity) curves (without temporal prior) are displayed in Fig.\,\ref{Fig:SSEcomparison} by adjusting the weight $\alpha$. The AUCs for JVSS and SSE are 0.9471 and 0.8985, respectively. When the precision is around 0.93, the sensitivity of SSE is 0.673 (187/278) while that of JVSS is 0.860 (239/278); when the sensitivity is around 0.88, the precision of SSE is 0.631 (246/(246 + 144)) while that of JVSS is 0.918 (245/(245 + 22)). Fig.\,\ref{Fig:SSEcomparison} demonstrates that our proposed JVSS loss has better metastasis detection accuracy than the SSE loss.

\begin{table}[h]
\begin{center}
\captionv{15}{}{Metastasis detection accuracy of 3D U-Net with different $\alpha$ values.}
\label{Tab:MetDetLossUNet}
\begin{tabular}{|l|l|c|c|c|c|c|c}
\Xhline{3\arrayrulewidth}
 Network & Loss & Sensitivity & Precision & $\#$FP & mDSC \\
\Xhline{3\arrayrulewidth}
\multirow{3}{*}{\modified{3D U-Net}} &BCE &0.720 & 0.667 & 100 & 0.754 \\
\cline{2-6}
& JVSS$_{\alpha =0.9}$ &0.827 &0.197 & 939 & 0.814\\
& JVSS$_{\alpha =0.5}$ &0.694 &0.835 & 38 & 0.763\\
\Xhline{3\arrayrulewidth}
\end{tabular}
\end{center}
\vspace{10pt}
\end{table}
Our proposed JVSS loss works with other neural networks. As an example, the metastasis detection accuracy of the JVSS loss with 3D U-Net \cite{hu2019multimodal} is displayed in Tab.\,\ref{Tab:MetDetLossUNet}. With the BCE loss, 3D U-Net achieves a sensitivity of 0.720 (200/278) and a precision of 0.667 (200/(200 + 100)), which is consistent with the performance reported by other groups \cite{hu2019multimodal}. With JVSS, when $\alpha = 0.9$, the sensitivity of 3D U-Net increases to 0.827 (230/278), but the precision drops to 0.197 {(230/(230 + 939)}; when $\alpha = 0.5$, the sensitivity is 0.694 {(193/278)} but the precision increases to 0.835 {(193/(193 + 38))}. However, the performance of 3D U-Net is inferior to DeepMedic in general. Therefore, in this work DeepMedic is chosen as our baseline method. Due to high computation of 3D models, only $\alpha = 0.9$ and $\alpha = 0.5$ are displayed for 3D U-Net as demonstration examples.

\subsection{Results of DeepMedic+ using temporal prior}
\begin{figure}
\centering
\begin{minipage}[b]{0.24\linewidth}
\subfigure[Prior image]{
\includegraphics[width=\linewidth]{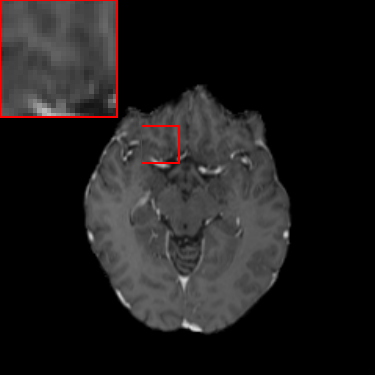}
\label{subfig:volumeS60VorPrior}
}
\end{minipage}
\begin{minipage}[b]{0.24\linewidth}
\subfigure[Main image with mask]{
\includegraphics[width=\linewidth]{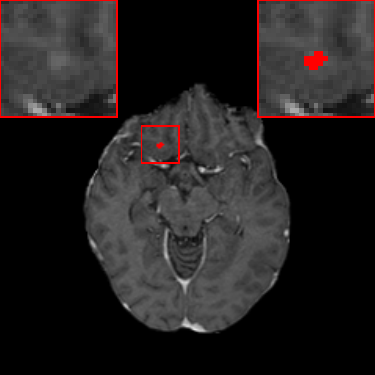}
\label{subfig:volumeS60VorRGB}
}
\end{minipage}
\begin{minipage}[b]{0.24\linewidth}
\subfigure[Difference ((b)-(a))]{
\includegraphics[width=\linewidth]{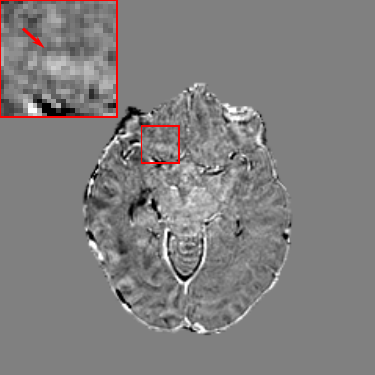}
\label{subfig:volumeS60VorDiffPrior}
}
\end{minipage}
\begin{minipage}[b]{0.24\linewidth}
\subfigure[Posterior image]{
\includegraphics[width=\linewidth]{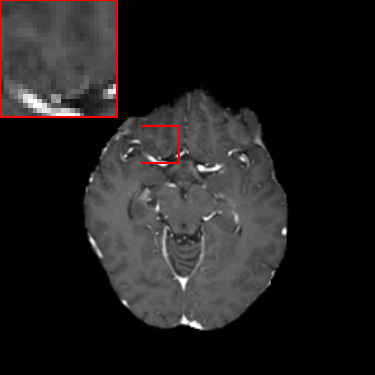}
\label{subfig:volumeS60VorPost}
}
\end{minipage}
\captionv{15}{}{The single false positive metastasis detected by $\mathcal{M}_\text{spec}$: (a) temporal prior image; (b) current main image, where the FP metastasis is marked by red color; (c) difference image ((b)-(a)); (d) temporal posterior image. }
\label{Fig:FalsePositivePriorCases}
\end{figure}

Fig.\,\ref{subfig:volumeS59RGB} is an example where temporal prior information is beneficial for metastasis identification. With an additional path for the prior volume, the red region in Fig.\,\ref{subfig:volumeS59RGB}, as well as the other two cases in Fig.\,\ref{Fig:FalsePositiveCases}, is detected correctly. The sensitivity and precision for {DeepMedic+} with the JVSS loss ($\alpha = 0.995$ and $\alpha = 0.5$) together with the temporal prior, i.e. $\mathcal{M}_\text{sens}$ and $\mathcal{M}_\text{spec}$ respectively, are displayed in Tab.\,\ref{Tab:MetDetLoss}. For $\alpha = 0.995$ ($\mathcal{M}_\text{sens}$), the total number of FP metastases is reduced from 247 to 158, where {$36.0\%$ ((247 - 158)/247)} FP metastases are reduced. {The average FP rate per patient is reduced from 2.40 (247/103) to 1.53 (158/103).} As a trade-off, the sensitivity has a slight decrease from 0.946 {(263/278)} to 0.932 {(259/278)}, with only 4 more FN metastases. Note that some of the test volumes are first scans without temporal prior volumes. If we exclude such volumes, the number of FP metastases decreases from 180 to 100 with the help of temporal prior, which is about {44.4\% (180 - 100)/180} less, where 3 instead of 4 more FN metastases are observed. {The average FP rate per patient is reduced from 2.54 (180/71) to 1.41 (100/71) for those with prior scans.} For $\alpha = 0.5$ ($\mathcal{M}_\text{spec}$), the sensitivity is slight worse than that of $\mathcal{M}_\text{baseline}$. However, the precision is as high as 0.996 {(234/(234 + 1))} with only one FP metastasis. The FP case is displayed in Fig.\,\ref{Fig:FalsePositivePriorCases}, where the current main image together with its temporal prior and posterior images are displayed. The difference image between the main image and the temporal prior image is displayed in Fig.\,\ref{subfig:volumeS60VorDiffPrior}, where the area indicated by the arrow has larger difference. That is why $\mathcal{M}_\text{spec}$ regards this region as a metastasis. However, after checking its posterior image (Fig.\,\ref{subfig:volumeS60VorPost}), no grown metastasis exists. Therefore, we regard the detection in Fig.\,\ref{subfig:volumeS60VorRGB} as FP. But we cannot eliminate the possibility that a real metastasis has regressed at the segmented region before the posterior scan.

\  

\begin{figure}
\centering
\begin{minipage}{0.3\linewidth}
\subfigure[0.941, 0.916]{
\includegraphics[width=\linewidth]{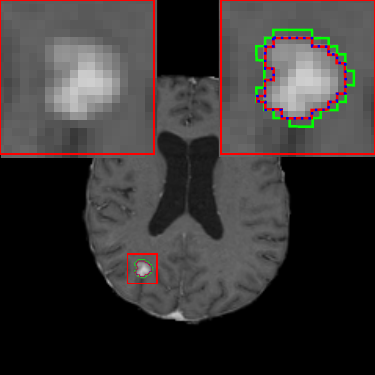}
\label{subfig:volumeS90DSC}
}
\end{minipage}
\begin{minipage}{0.3\linewidth}
\subfigure[0.910, 0.910]{
\includegraphics[width=\linewidth]{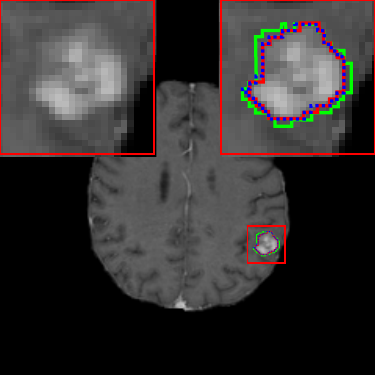}
\label{subfig:volumeS102DSC}
}
\end{minipage}
\begin{minipage}{0.3\linewidth}
\subfigure[0.596, 0.133]{
\includegraphics[width=\linewidth]{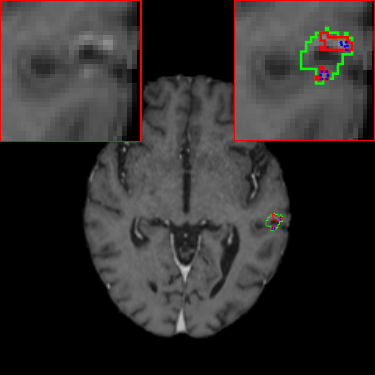}
\label{subfig:volumeS67DSC}
}
\end{minipage}
\captionv{15}{}{Segmentation boundaries of three exemplary metastases. Green boundaries are manual reference segmentation boundaries, while {red solid and blue dotted} boundaries are segmentation boundaries of $\mathcal{M}_\text{sens}$ and $\mathcal{M}_\text{spec}$, respectively. The DSC values of each metastasis volume are displayed on the left and right in the subscaptions for $\mathcal{M}_\text{sens}$ and $\mathcal{M}_\text{spec}$, respectively. {Zoom in for better visualization.}}
\label{Fig:SegmentationBoundaries}
\end{figure}

For segmentation accuracy, the mDSCs for all the TP metastases are displayed in Tab.\,\ref{Tab:MetDetLoss}. In general, the mDSCs are around 0.8. With a smaller $\alpha$ value, the mDSC value gets slightly smaller. The segmentation boundaries of three exemplary metastases are displayed in Fig.\,\ref{Fig:SegmentationBoundaries}. Green boundaries are manual reference segmentation boundaries, while red ($\alpha = 0.995$) and blue ($\alpha = 0.5$) boundaries are segmentation boundaries of $\mathcal{M}_\text{sens}$ and $\mathcal{M}_\text{spec}$, respectively. In Figs.\,\ref{Fig:SegmentationBoundaries}(a) and (b), all the three boundaries have good consistency, with DSC values larger than 0.9. These two segmentation results are general cases for metastases larger than 0.1\,cm$^3$. Fig.\,\ref{subfig:volumeS67DSC} is one example where {DeepMedic+} achieves lower DSC values. The tumorous region in this case is difficult to define in {single-sequence} data as some part of the whole tumor has necrosis after treatment. In addition, the active parts, which are enhanced by contrast agents, are separated in many {axial} slices. Therefore, $\mathcal{M}_\text{sens}$ and $\mathcal{M}_\text{spec}$ both segment the active parts as two separated metastases. As a consequence, the DSC values are low. Nevertheless, $\mathcal{M}_\text{sens}$ segments the active parts better than $\mathcal{M}_\text{spec}$. Tab.\,\ref{Tab:MetDetLoss} indicates that $\mathcal{M}_\text{sens}$ achieves better mDSC values than the baseline DeepMedic, while $\mathcal{M}_\text{spec}$ is slightly worse. Since the union of $\mathcal{M}_\text{sens}$ and $\mathcal{M}_\text{spec}$ segmentation masks is used in our ensemble learning, the good segmentation of $\mathcal{M}_\text{sens}$ is preserved.

\subsection{{DeepMedic+} ensemble}
\begin{figure}
\centering
\begin{minipage}{0.3\linewidth}
\centering
Reference
\end{minipage}
\begin{minipage}{0.3\linewidth}
\centering
Baseline
\end{minipage}
\begin{minipage}{0.3\linewidth}
\centering
Proposed
\end{minipage}

\begin{minipage}{0.3\linewidth}
\subfigure[]{
\includegraphics[width=\linewidth]{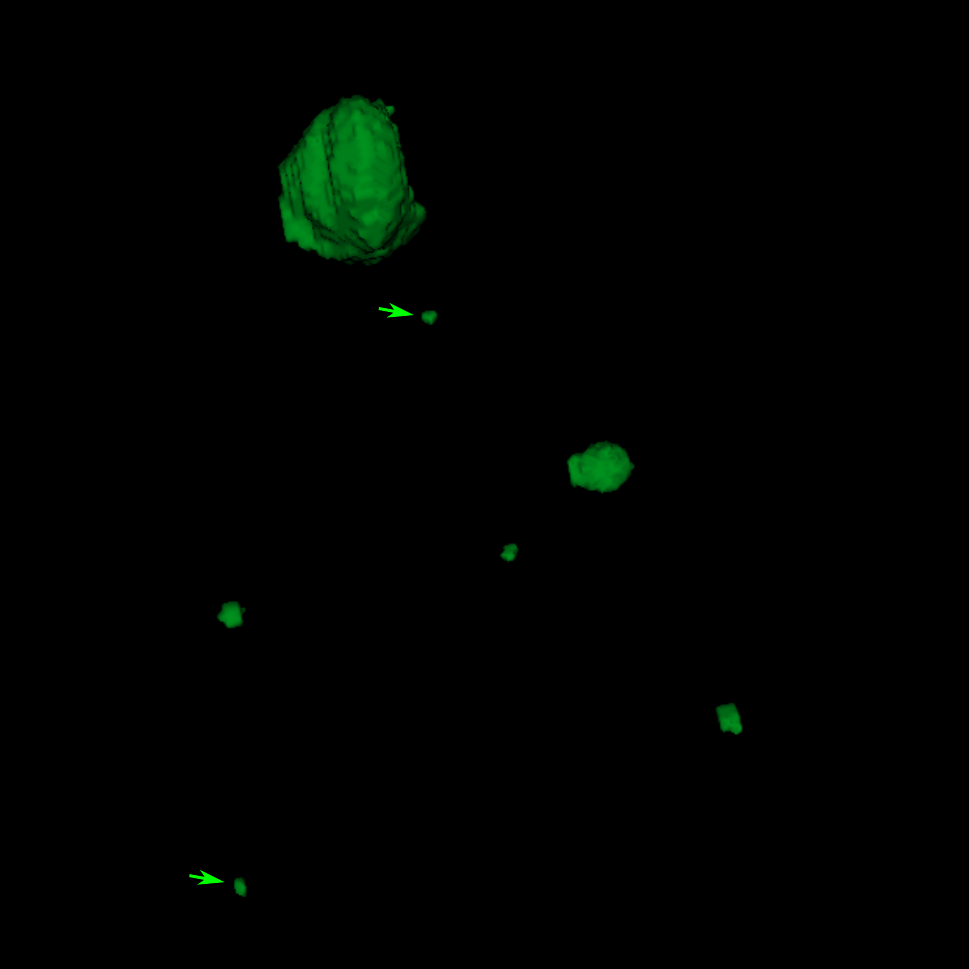}
\label{subfig:ensembleRef1}
}
\end{minipage}
\begin{minipage}{0.3\linewidth}
\subfigure[]{
\includegraphics[width=\linewidth]{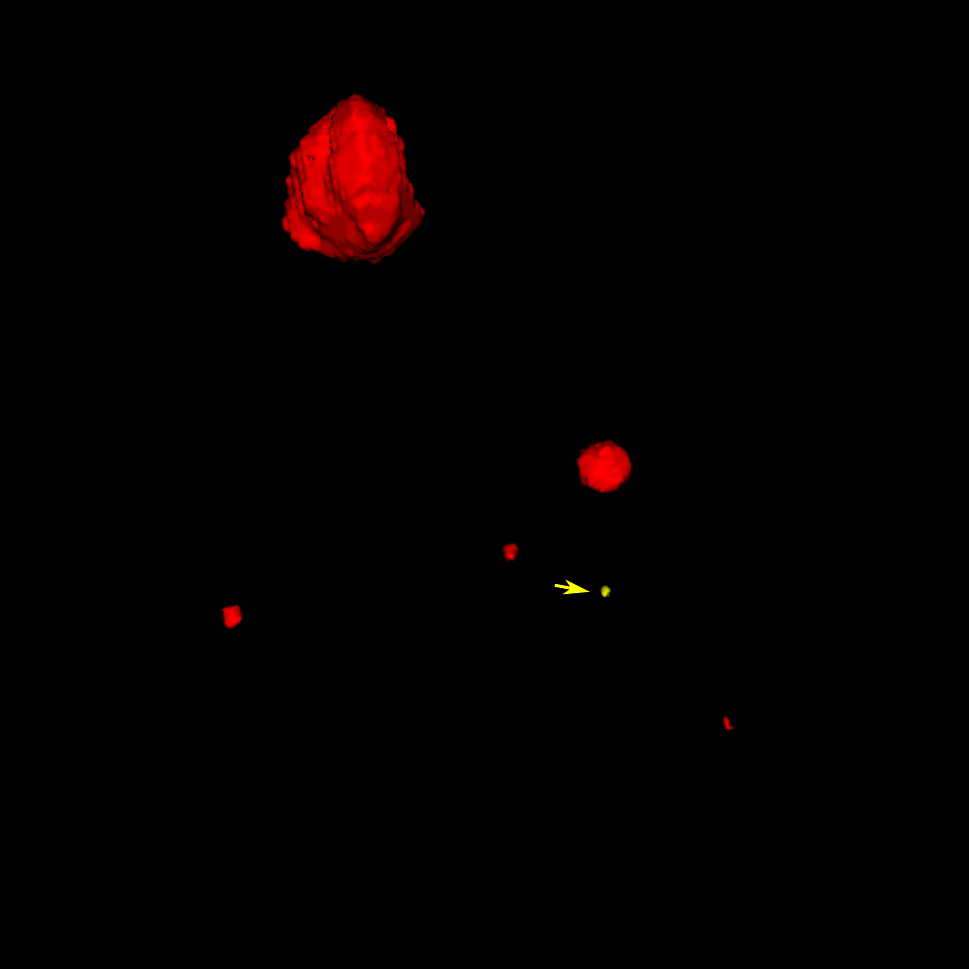}
\label{subfig:ensembleRegular1}
}
\end{minipage}
\begin{minipage}{0.3\linewidth}
\subfigure[]{
\includegraphics[width=\linewidth]{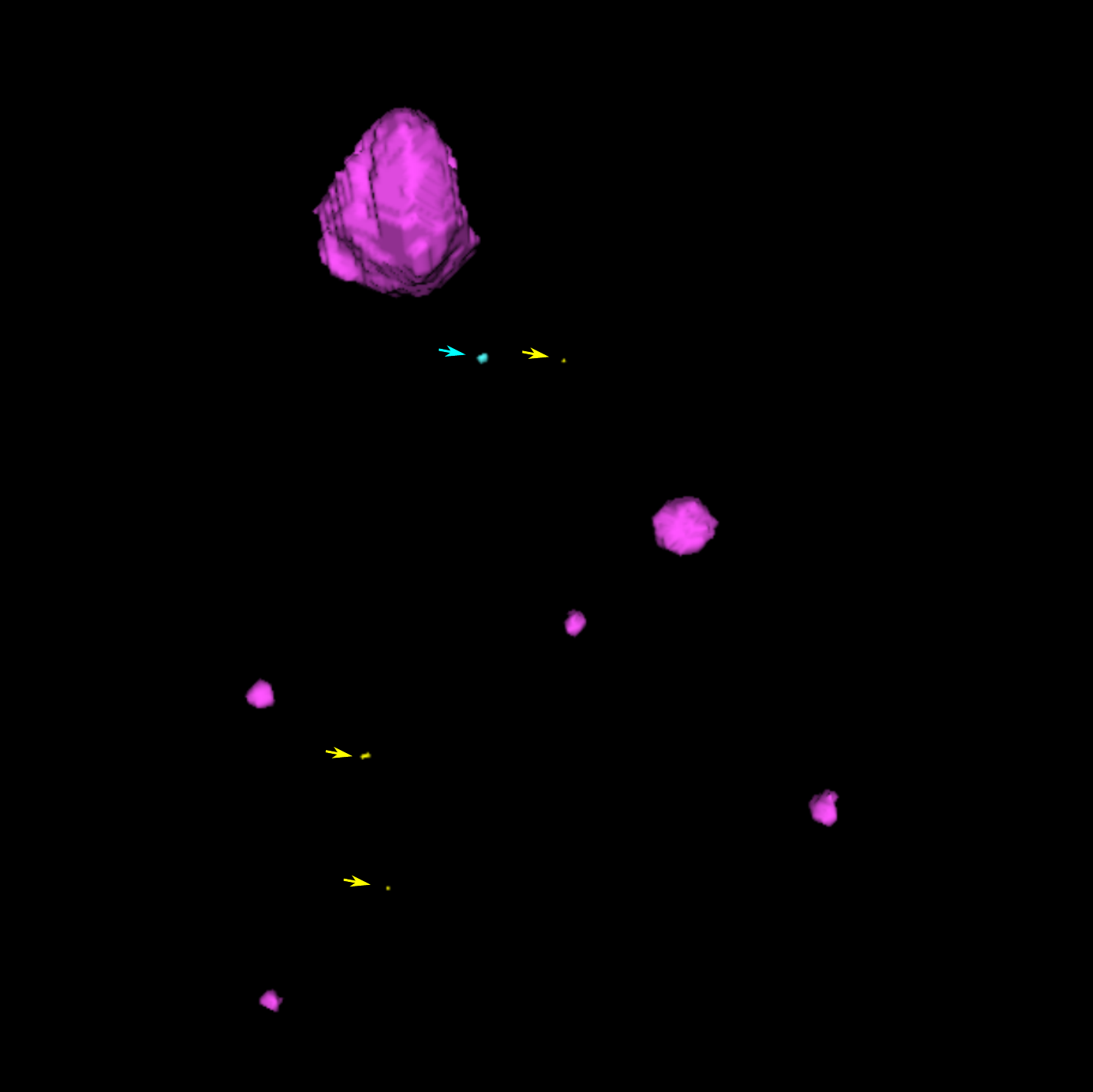}
\label{subfig:ensemble1}
}
\end{minipage}

\begin{minipage}{0.3\linewidth}
\subfigure[]{
\includegraphics[width=\linewidth]{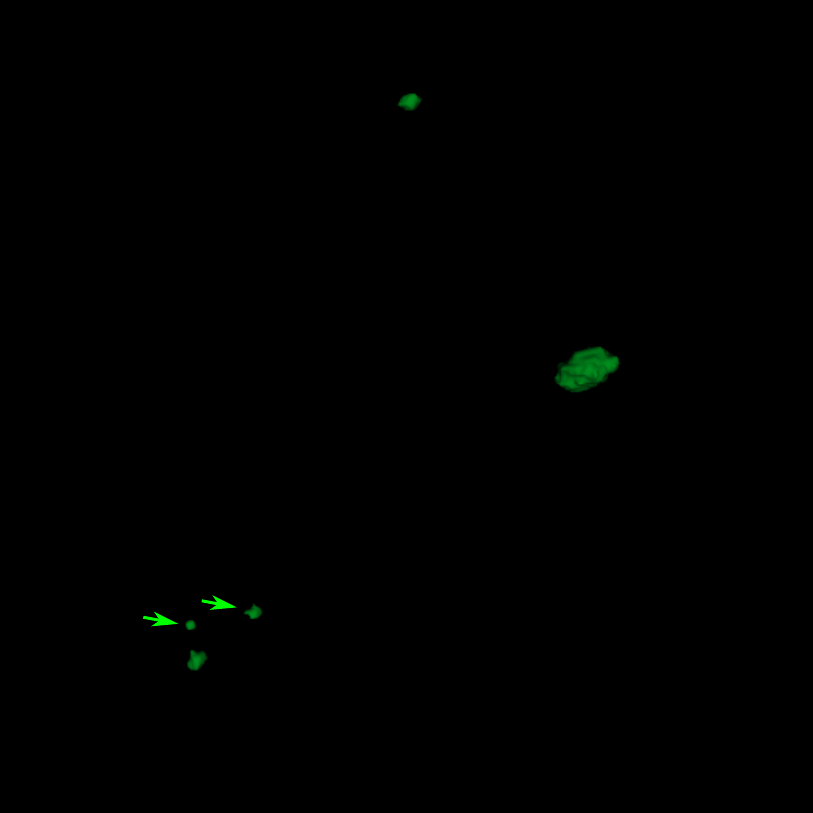}
\label{subfig:ensembleRef2}
}
\end{minipage}
\begin{minipage}{0.3\linewidth}
\subfigure[]{
\includegraphics[width=\linewidth]{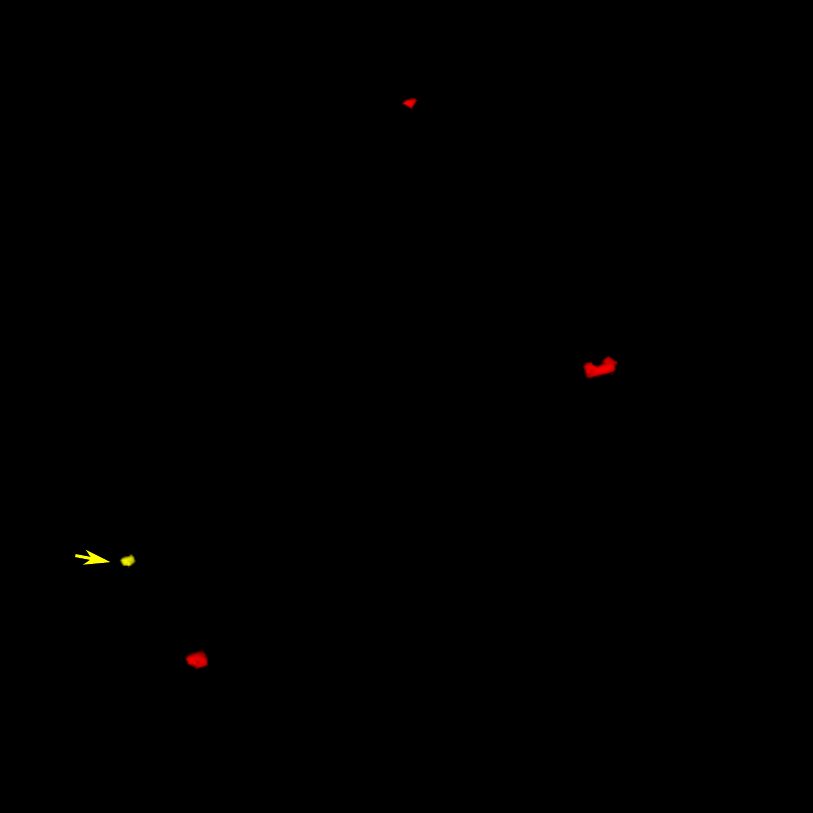}
\label{subfig:ensembleRegular2}
}
\end{minipage}
\begin{minipage}{0.3\linewidth}
\subfigure[]{
\includegraphics[width=\linewidth]{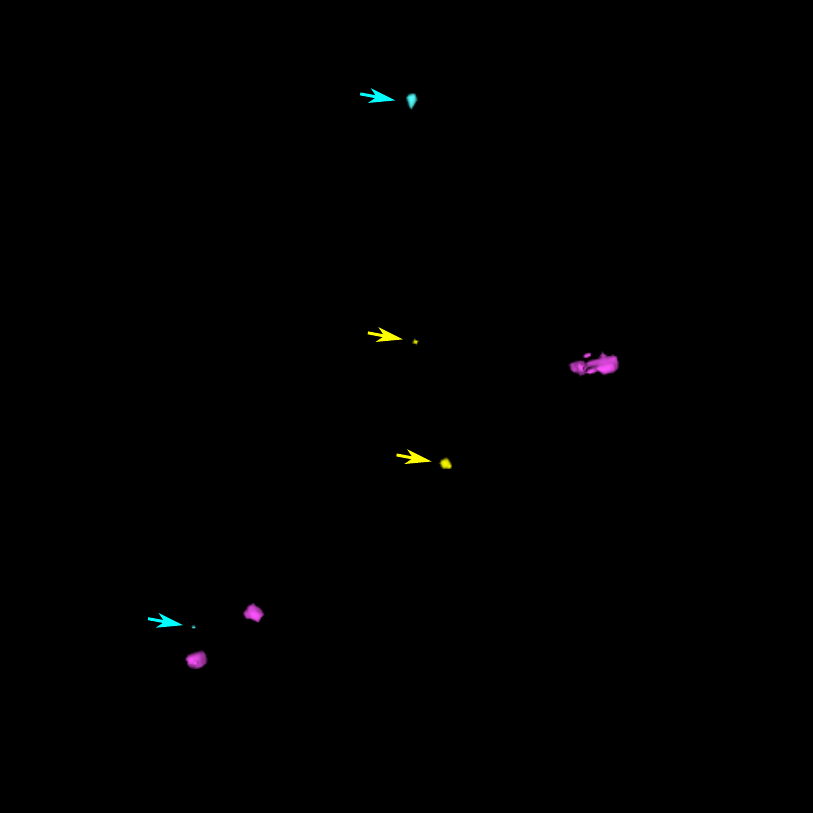}
\label{subfig:ensemble2}
}
\end{minipage}

\begin{minipage}{0.3\linewidth}
\subfigure[]{
\includegraphics[width=\linewidth]{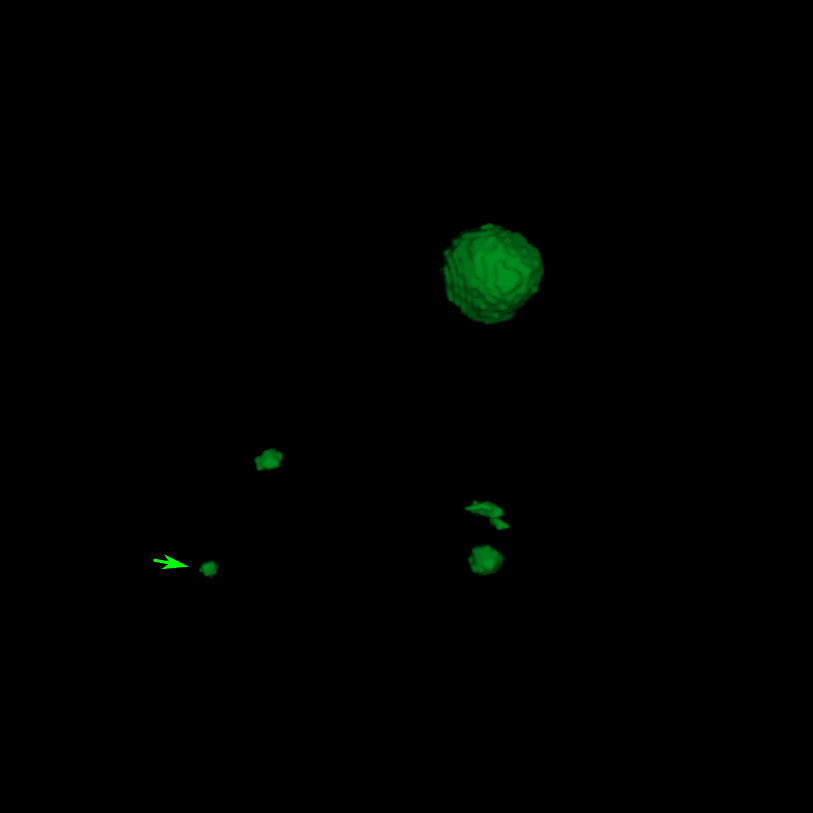}
\label{subfig:ensembleRef3}
}
\end{minipage}
\begin{minipage}{0.3\linewidth}
\subfigure[]{
\includegraphics[width=\linewidth]{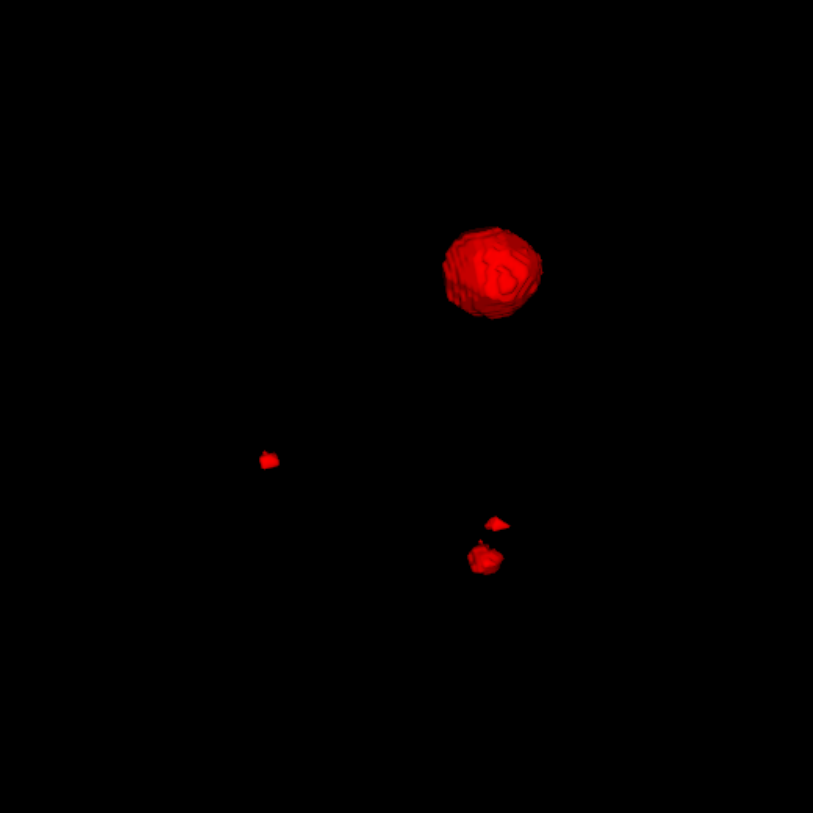}
\label{subfig:ensembleRegular3}
}
\end{minipage}
\begin{minipage}{0.3\linewidth}
\subfigure[]{
\includegraphics[width=\linewidth]{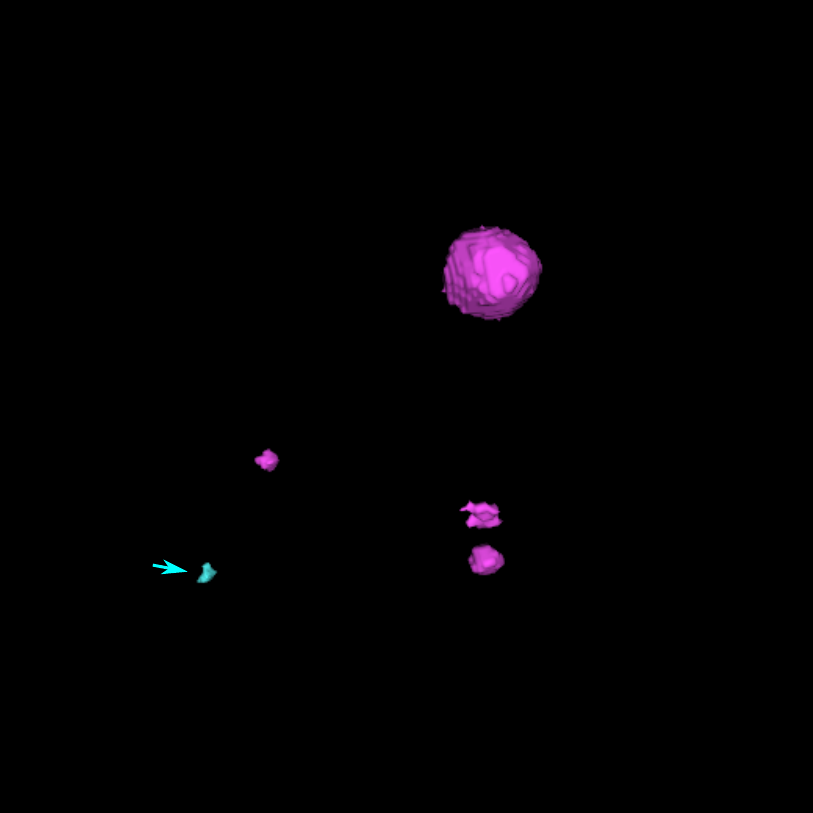}
\label{subfig:ensemble3}
}
\end{minipage}

\captionv{15}{}{The segmentation results of three exemplary patients (Patient 1: (a)-(c); Patient 2: (d)-(f); Patient 3: (g)-(i)) by the baseline DeepMedic $\mathcal{M}_\text{baseline}$ and {DeepMedic+} ensemble ($\mathcal{M}_\text{spec}$ + $\mathcal{M}_\text{sens}$), rendered by 3D Slicer. Left: ground truth (green), middle: prediction of original DeepMedic, right: our proposed {DeepMedic+} ensemble. The metastases indicated by the green arrow are missed by $\mathcal{M}_\text{baseline}$, but detected by DeepMedic ensemble. The red metastases are TP metastases detected by baseline DeepMedic. The pink metastases are true positive metastases detected by $\mathcal{M}_\text{spec}$. The cyan metastases indicated by cyan arrows are additional true positive metastases detected by $\mathcal{M}_\text{sens}$. The yellow metastases indicated by yellow arrows are false positive metastases. }
\label{Fig:ensemble}
\end{figure}

The segmentation results of three exemplary patients by the baseline DeepMedic $\mathcal{M}_\text{baseline}$ and {DeepMedic+} ensemble ($\mathcal{M}_\text{spec}$ + $\mathcal{M}_\text{sens}$), rendered by 3D Slicer, are displayed in Fig.\,\ref{Fig:ensemble}. The metastases indicated by the green arrows are missed by $\mathcal{M}_\text{baseline}$, but these missing ones are all detected by DeepMedic ensemble. The red metastases are TP metastases detected by $\mathcal{M}_\text{baseline}$. The pink metastases are detected by $\mathcal{M}_\text{spec}$, which are all TP. The cyan metastases are additional TP metastases detected by $\mathcal{M}_\text{sens}$. In Fig.\,\ref{subfig:ensemble1} and Fig.\,\ref{subfig:ensemble3}, one additional TP metastasis is detected by $\mathcal{M}_\text{sens}$, while two additional TP metastases are detected in Fig.\,\ref{subfig:ensemble2}. The yellow metastases are FP metastases. One FP metastasis is present in Fig.\,\ref{subfig:ensembleRegular1} and Fig.\,\ref{subfig:ensembleRegular2}. In Fig.\,\ref{subfig:ensemble1} and Fig.\,\ref{subfig:ensemble2}, three and two FP metastases are present, respectively. Fig.\,\ref{Fig:ensemble} demonstrates that with the ensemble of $\mathcal{M}_\text{spec}$ and $\mathcal{M}_\text{sens}$, the majority of TP metastases are confirmed. Only a few metastasis candidates marked exclusively by $\mathcal{M}_\text{sens}$ in every patient need expert confirmation, additional diagnostic studies or imaging follow-up. Note that the three examples in Fig.\,\ref{Fig:ensemble} have relatively a large number of metastases. On average, each patient volume has about {0.24 ((259 - 234)/103)} additional TP metastases to include and {1.52 ((158 - 1)/103)} additional FP metastases to exclude, according to Tab.\,\ref{Tab:MetDetLoss}. {Among the additional 25 TP metastases, 21 are smaller than $0.1\,\textrm{cm}^3$, which is 84\% (21/25). Among the 157 additional FP metastases, 154 are smaller than $0.1\,\textrm{cm}^3$, which is 98.1\% (154/157).}

\section{Discussion}
With the selected subvolume size $37^3$, each subvolume typically contains a single metastasis or no metastasis, so that the sensitivity and specificity definitions derived from predicted voxel-level probabilities in the JVSS loss should impact the sensitivity and specificity for detection of individual metastases. Although a few subvolumes may contain more than one metastases in the training data, they do not make a significant difference due to their low incidence. For inference, the trained model is still able to detect multiple metastases in a single subvolume, as demonstrated in Fig.\,\ref{subfig:ensemble1} where two close metastases are both detected by $\mathcal{M}_\text{sens}$. {Since the used DeepMedic network has a receptive field of $19^3$, our JVSS loss requires the input subvolume size larger than $19^3$. A slight different subvolume size from $37^3$, for example $30^3$, has no big impact on the accuracy theoretically, but the training efficiency will change.}

{Tab.\,\ref{Tab:MetDetLoss} shows that the proposed JVSS loss (e.g., $\alpha = 0.95$) can improve brain metastasis identification. This indicates that the two losses BCE and VSS in Eqn.\,(\ref{eqn:MetDetLoss}) are not conflicting objectives; otherwise, the performance will be worse. Different weights can also be used to combine these two losses in Eqn.\,(\ref{eqn:MetDetLoss}). They will lead to different performances. On the one hand, it is computationally expensive to find such an optimal weight parameter. On the other hand, comparable performance (likely not the exactly same performance) can also be achieved by tuning the $\alpha$ value. For example, $\ell_{\text{JVSS}} = \frac{2}{3}\ell_{\text{VSS}} + \frac{1}{3} \ell_{\text{BCE}}$ or $\ell_{\text{JVSS}} = \frac{1}{3}\ell_{\text{VSS}} + \frac{2}{3} \ell_{\text{BCE}}$ can also be used. The former has a weight ratio of 2.0, while the latter has a weight ratio of 0.5. The ratio 2 achieves 0.906, 0.750, 84, and 0.786 for metastasis-level sensitivity, precision, $\#$FP and mDSC, respectively, when $\alpha$ is fixed to 0.95. The performance metrics can be achieved by selecting another $\alpha$ value between 0.99 and 0.95 using $\ell_{\text{JVSS}} = \ell_{\text{VSS}} + \ell_{\text{BCE}}$. Similarly, the ratio 0.5 achieves 0.906, 0.708, 104 and 0.774 for metastasis-level sensitivity, precision, $\#$FP and mDSC, respectively. Therefore, in this work we choose a fixed weight to avoid new hyper-parameters.}

In Fig.\,\ref{Fig:FalseNegativeExamples}, the volume sizes of the three metastases are all smaller than 0.1\,cm$^3$. Therefore, it is very challenging for DeepMedic, and for human experts as well, to identify them, especially with many vascular structures being similar to them. These small metastases consist of very few voxels and there is ambiguity due to partial volume effects in the periphery of the metastases. For such tiny metastases, DeepMedic only detects a few voxels, or even one voxel only, in each metastasis region. For example, in Fig.\,\ref{subfig:volumeS36RGB}, not all the metastasis voxels are covered by the segmentation mask. Nevertheless, detecting their existence already has important clinical value. In general the segmentation accuracy of DeepMedic does not change or changes slightly for all other large metastases. This is demonstrated by the mDSC values in Tab.\,\ref{Tab:MetDetLoss} and the 3D rendering in Fig.\,\ref{Fig:ensemble}. Interestingly, mDSC is even improved using the proposed JVSS loss in $\mathcal{M}_\text{sens}$ as compared to BCE alone, which illustrates that the introduction of the proposed VSS loss does not compromise the accuracy of individual segmentation masks in detected metastases.


Ensemble learning is also used by other research groups for BM identification \cite{hu2019multimodal,bousabarah2020deep}. Hu et al. \cite{hu2019multimodal} proposed to use the average probability map from U-Net and DeepMedic. However, for each individual segmented metastasis, there is still no guarantee whether it is a TP or FP metastasis. Bousabarah et al. \cite{bousabarah2020deep} applied majority voting to decide the final segmentation mask from three different U-Net models to achieve high precision. However, its sensitivity is only 0.77. Our ensemble learning puts high sensitivity and high precision together in one result. As displayed in Fig.\,\ref{Fig:ensemble} and Tab.\,\ref{Tab:MetDetLoss}, the metastases detected by $\mathcal{M}_\text{spec}$ are 99.6\% (almost 100\%) sure to be TP metastases. For the remaining metastasis candidates detected by $\mathcal{M}_\text{sens}$, it is very efficient to further evaluate whether they are TP or FP as there are only 1.75 metastases per patient remaining on average. Compared with the state-of-the-art methods, the FP rate of our method is very small, {because of the used temporal prior}. For example, the DeepMedic method proposed by Charron et al. \cite{charron2018automatic} achieves a sensitivity of 93\%, but with 7.8 FP metastases per patient. The ensemble solution thereby combines high confidence for metastasis detection with high sensitivity, efficiently guiding expert attention to metastases candidates that are uncertain or require additional follow-up, where it is required the most. We therefore believe that this approach is especially well-fit to the requirements of expert support in real clinical practice.

Note that DeepMedic is an exemplary network to demonstrate the efficacy of our proposed sensitivity-specificity loss and temporal prior for BM identification in longitudinal MRI data. We do not claim DeepMedic the best network for BM identification. With the rapid development of deep learning techniques, new neural architectures or existing neural architecture variants with our proposed VSS loss and temporal prior may achieve comparable or even better performance. Nevertheless, the ensemble approach of {DeepMedic+} has already important value in real clinical practice.

{In this work, a post-contrast T1-MPRAGE sequence was used for imaging of BM. T1-MPRAGE is a standard 3D T1 inversion recovery gradient echo sequence (3D T1 IR-GRE) with MPRAGE being the Siemens/Hitachi implementation and IR-SPGR/Fast SPGR with inversion activated or BRAVO being the GE version, 3D turbo field echo (TFE) being the Philips and 3D Fast FE being the Toshiba implementation of a standard 3D T1 IR-GRE sequence. 3D T1 IR-GRE sequences are standard 3D T1w MR sequences for BM imaging. In fact, the 2020 consensus brain tumor imaging protocol for clinical trials in BM \cite{kaufmann2020consensus} recommends a 3D T1w IR-GRE sequence as standard post-contrast 3D T1w sequence in their ``Minimum standard 1.5T metastatic brain tumor imaging protocol" and their ``Minimum standard 3T metastatic brain tumor imaging protocol". An alternative 3D T1w sequence option for BM are 3D T1 TSE sequences. 3D T1 IR-GRE sequences have increased contrast between grey and white matter and more pronounced visualization of vessels compared to 3D T1 Turbo spin-echo (TSE) sequences \cite{danieli2019brain}. Because of this characteristic, we would expect automatic segmentation of BM in T1 IR-GRE to be more challenging for a 3D-CNN compared to T1 TSE sequences. However, for older T1 weighted echo-spin (ES) sequence, due to the lower anatomic contrast compared with MPRAGE and coarser resolution in the axial direction \cite{brant1992mp}, a larger false negative rate for small metastases is expected. Comparison of model performance for different sequence types and defining optimal sequences for deep learning-based metastases detection is an interesting research question, which should be pursued in future work. Irrespective of the MRI sequence, images of the current dataset were characterized by a high signal-to-noise ratio and all the volumes with apparent motion artifacts were excluded. The performance of our algorithm at lower signal-to-noise ratios (low field strength, short acquisition times, poor receiver coil characteristics etc.) and motion-corrupted images also needs further evaluation.}

\section{Conclusion And Outlook}
Tiny BM detection is a challenging task due to their small size, low contrast, and similar appearance to vessels. Our proposed JVSS loss can adjust the detection sensitivity and precision in a large range to either achieve high sensitivity or high precision. The temporal prior information is able to further reduce FP metastases. The ensemble learning is able to distinguish high confidence true positive metastases from metastases candidates that require special expert attention or further follow-up, which facilitates metastasis detection and segmentation for neuroradiologists in diagnostic and radiation oncologists in therapeutic clinical applications.

\section*{References}
\addcontentsline{toc}{section}{\numberline{}References}
\vspace*{-10mm}










\end{document}